\documentclass[preprint,1p]{elsarticle}
\usepackage[T1]{fontenc}
\usepackage[utf8]{inputenc}
\usepackage{lmodern, graphicx, amsmath, physics, booktabs, textcomp}
\usepackage{siunitx}
\usepackage[version=4]{mhchem} 
\usepackage{float}
\usepackage{hyperref}
\usepackage[nameinlink, capitalize]{cleveref}

\makeatletter
\def\ps@pprintTitle{%
 \let\@oddhead\@empty
 \let\@evenhead\@empty
 \def\@oddfoot{}%
 \let\@evenfoot\@oddfoot}
\makeatother

\begin{document}

\begin{frontmatter}

\title{Assessment of mechanical, thermal properties and crystal shapes of 
monoclinic tricalcium silicate from atomistic simulations}

\author[add1,add2]{Jérôme Claverie}
\author[add1]{Siham Kamali-Bernard}
\ead{siham.kamali-bernard@insa-rennes.fr}
\author[add2]{João Manuel Marques Cordeiro}
\author[add1]{Fabrice Bernard}

\address[add1]{Laboratory of Civil Engineering and Mechanical Engineering
				(LGCGM), INSA Rennes, Rennes, France} \address[add2]{Department
				of Physics and Chemistry, School of Natural Sciences and
				Engineering, São Paulo State University (UNESP), 15385-000 Ilha
				Solteira, São Paulo, Brazil}

\begin{abstract}
  The two most common polymorphs in industrial alite, M1 and M3, were
  characterized at the molecular scale. Different methods were employed and
  discussed to assess mechanical properties and specific heat of both
  polymorphs. The calculated homogenized elastic moduli and specific heat were
  found in good agreement with experimental measurements. A comparative analysis
  of spacial Young's modulus reveal isotropic and anisotropic spacial
  distribution for \ce{M1} and \ce{M3} respectively. A more isotropic
  compressive strength is also reported for \ce{M1} when compared to \ce{M3}
  polymorph. Cleavage energies computation allowed to proposed equilibrium
  shapes for both polymorph, with significant differences. While the lowest
  cleavage energies were found along (100) and (001) for both polymorphs, the
  constructed M1 crystal possesses 3 independent facets, against seven for the
  M3 polymorph.
\end{abstract}


\end{frontmatter}

\section{Introduction}

The research in cementitious materials is experiencing new challenges mainly due to the need to preserve the environment and save energy. To reduce \ce{CO2} emissions and energy cost of production, alternative binders are under study \cite{Biernacki2017}. However, ordinary Portland cement (OPC) should continue to be employed for a long time and understanding of its principal constituents is primordial for its improvement. Another way to reduce the environmental impact of Portland cement is to enhance its reactivity which will involve less content of cement in concrete for the same strength. Since alite is the principal phase of Portland clinker that most contributes to strength development of Portland cement and particularly at early ages, a deep understanding of its properties is crucial to improve its quality and reactivity. Alite is a tricalcium silicate (\ce{C3S}) with minor oxides usually called impurities. It presents a large grade of polymorphism depending on different factors, among them: the nature and the amount of impurities, the temperature of preheating or burning \cite{Stanek2002}. Seven structures were reported in industrial alite: three triclinic (\ce{T1}, \ce{T2} and \ce{T3}) , three monoclinic (\ce{M1}, \ce{M2} and \ce{M3}), and a rhombohedral form R. These polymorphs appear via successive and reversible phase transitions \cite{Taylor1990}:
\begin{equation}
  \ce{T1 <->[\SI{620}{\celsius}] T2 <->[\SI{920}{\celsius}] T3 <->[\SI{980}{\celsius}] M1 <->[\SI{990}{\celsius}] M2 <->[\SI{1060}{\celsius}] M3 <->[\SI{1070}{\celsius}] R}
\end{equation}
It is well known that impurities in alite stabilize high temperature polymorphs
at low temperatures \cite{Taylor1990, Maki1984}. The two main
\ce{C3S} polymorphs present in industrial clinker are \ce{M1} and \ce{M3}. According to Maki and Goto \cite{Maki1982a}, \ce{MgO} in clinker promotes the
stable growth of alite in favor of the occurrence of \ce{M3}. In contrary,
decreased \ce{MgO}/\ce{SO3} ratio lead to \ce{M1} stabilization
\cite{Maki1982a,Stanek2002}. It was also reported that preheating of the raw
meal may result in the disappearance of the M3 polymorph \cite{Stanek2002}. This
modifications in the structure of \ce{C3S} can have a significant impact on
strength as reported in few experimental results in the literature
\cite{Stanek2002, Zhou2018}.  The transformation of \ce{M3} to \ce{M1}
polymorph may result in a 10\% increase in the compressive strength
\cite{Stanek2002}. The origin of this observed variation in strength could be
explained by a greater amount of non-bonding electrons in oxygens of \ce{M1
C3S}, leading to a higher reactivity when compared to the \ce{M3} polymorph. The
impact of structure at the nanoscale on the properties like strength is a
complex topic and investigation at the atomic level via molecular modelling and
simulation should be helpful to improve our understanding of Portland cement. 

Over recent years, the properties of cementitious materials were addressed using
atomistic models, with particular attention on the main hydration product of
OPC: calcium silicate hydrates (C-S-H) \cite{Qomi2014a, Bauchy2014}. In
comparison, only few studies at the atomic scale focused on OPC clinker phases
\cite{Mishra2017}. Computation of thermal and mechanical properties at the
molecular scale can provides important information on the behaviour of OPC
clinker and hydrated product.  From the computation of surface energies, crystal
shapes can be theoretically constructed for different polymorphs and help to
understand morphological changes \cite{Azevedo2020}. The knowledge of
preferential cleavage planes and crystal shapes of \ce{C3S} polymorphs is
fundamental to understand their growth during the clinkering process
\cite{Maki1982}. It could also explain \ce{C3S} dissolution mechanisms
\cite{Nicoleau2015} or cleavage modes during clinker grinding \cite{Mishra2013,
Mishra2015}. Determination of such properties by experimental methods are most
of the time limited, especially in the case of surface energies \cite{Tran2016}.
In all cases, a proper synthesis procedure of pure \ce{C3S} is necessary, and
the determination of the amount of each polymorph in a sample is neither trivial
nor accurate \cite{Pustovgar2017}. 

In this work, the mechanical, thermal and surface properties of \ce{M1} and
\ce{M3 C3S} (the main forms of alite encountered in industrial OPC
\cite{Noirfontaine2012}) were characterized by molecular dynamics (MD)
simulations. 
\begin{figure}[!htbp]
  \centering
  \includegraphics[width=0.8\textwidth]{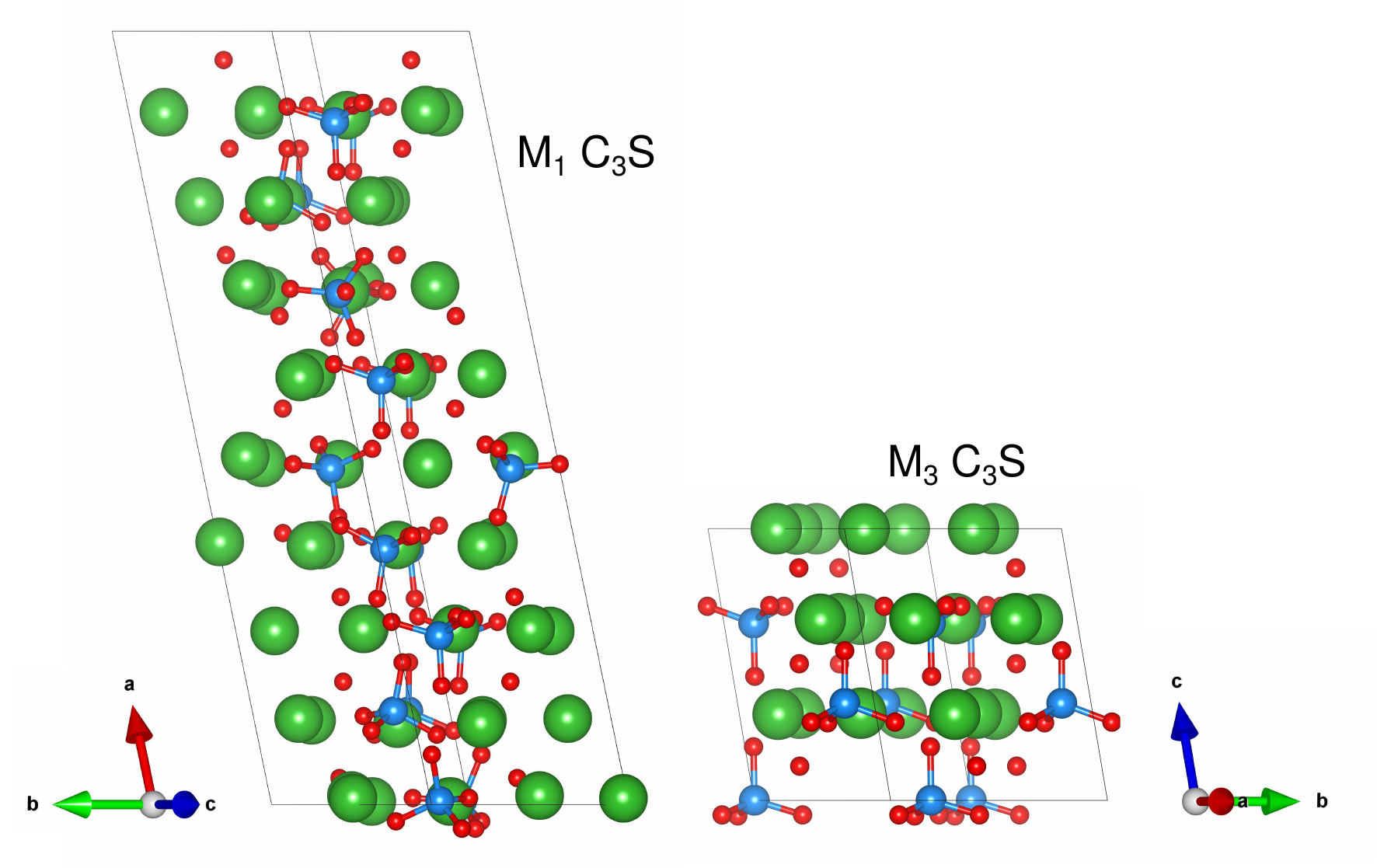}
  \caption{Unit cells of \ce{M1 C3S} (cell parameters: a =
  \SI{27.874}{\angstrom}, b = \SI{7.059}{\angstrom}, c = \SI{12.257}{\angstrom},
  $\beta = \ang{116.03}$) \cite{Noirfontaine2012} and \ce{M3 C3S} (cell
  parameters: a = \SI{12.235}{\angstrom}, b = \SI{7.073}{\angstrom}, c =
  \SI{9.298}{\angstrom}, $\beta = \ang{116.31}$) \cite{Mumme1995}.  Color code:
  calcium cations in green, oxygen  anions and silicate oxygen in red and silicon atoms in blue}
  \label{fig:unitcells}
\end{figure}
The present article is divided into four sections: crystal structures and force
fields, mechanical properties, thermal properties, and cleavage energies and
equilibrium shapes. The last three sections include a description of the method
employed and a presentation and discussion of results. To finish, a general
conclusion resumes the different findings.

\section{Crystal structures and force fields}

The atomistic systems investigated were built from the pure \ce{M1}
\cite{Noirfontaine2012} and \ce{M3} \cite{Mumme1995} crystal structures depicted
in \cref{fig:unitcells}. While the latter has already been used
\cite{Manzano2015, Mishra2013, Manzano2011}, this is the first time that a
\ce{M1 C3S} model has been employed in a MD investigation, despite of its
predominance in alite of Portland clinker with high \ce{SO3} content
\cite{Maki1982a}. Regarding atomic structural organization along the (010)
direction, and $b$ parameters, the two cells are very close. However, the two
models are shifted by 1/4 in cell units in the (010) direction, meaning that the
(010) Ca-rich plane of the \ce{M1} model corresponds to the (040) plane of the
\ce{M3} model. The atomic organization along (001) axis in the \ce{M1} model is
close to \ce{M3} model in the (100), and the $c$ and $a$ parameters in \ce{M1}
and \ce{M3} respectively are almost equal. Conversely, the $a$ parameter of the
\ce{M1} unit cell is approximately 3 times larger than the $c$ parameter for
\ce{M3}, and the major structural difference is expected in the (100) direction
for \ce{M1} and (001) direction for \ce{M3}.

Understanding the mechanical properties of \ce{C3S} is important for various reasons; among them: 1) the total hydration of a cement paste is never achieved, so clinker components are, to some extent, involved in the final microstructure of hydrated cement \cite{Ash1993, Vandamme2010}, and mainly at early ages; and 2) to optimize the grinding of clinker during cement manufacturing. The investigation of elastic properties of cementitious materials are most of the time related to hydrated products and very few data can be found in the literature concerning elastic properties of clinker components. Synthetic alite can be made by solid state sintering of decarbonated calcium oxide and fine silica, with possible addition of impurities (alumina, magnesium, sulfates), depending on the polymorph to be reached \cite{Nicoleau2013}. The elastic properties are typically determined by nanoindentation experiments and at the macroscale by resonance frequency measurements \cite{Velez2001,Brunarski1969}. Nanoindentation experiments are most of the time performed on hydrated mortar or cement paste \cite{Gao2017,Ulm2007,Vandamme2010,Puertas2011}, and more rarely on pure and doped clinker phases \cite{Velez2001}. Unhydrated clinker phases are known to exhibit stiffnesses by \numrange{3}{5} times larger and hardnesses by one order of magnitude larger than hydrated phases \cite{Velez2001,Vandamme2010,Manzano2009}.

Two force fields (FF) introduced in the \emph{cemff} database \cite{Mishra2017} were employed to describe atomic interactions in \ce{C3S}: INTERFACE FF (IFF) \cite{Heinz2013, Mishra2013} and ClayFF \cite{Cygan2004}. The PCFF implementation of IFF, used inhere, includes quadratic bonded terms for covalent bonds in silicates, an electrostatic term and a 9-6 Lennard-Jones (LJ) potential for short-range interactions:

\begin{multline} E_{IFF}=
\sum_{ij}\sum_{n=2}^{4}K_{r,ij}(r_{ij}-r_{0,ij})^n+
\sum_{ijk}\sum_{n=2}^{4}K_{\theta,ij}(\theta_{ij}-\theta_{0,ij})^n\\
+\frac{1}{4\pi\varepsilon_{0}\varepsilon_{r}}\sum_{ij}\frac{q_{i}q_{j}}{r_{ij}} +\sum_{ij}\varepsilon_{0,ij}\left[2\left(\frac{\sigma_{ij}}{r_{ij}}\right)^{9}- 3\left(\frac{\sigma_{ij}}{r_{ij}}\right)^{6}\right]
\label{eq:potential_iff}
\end{multline}

ClayFF is a general force field initially developed to describe interfaces
between clays and water \cite{Cygan2004}, and in particular adsorption of ions
for, inter alia, environmental application. It uses a flexible single point
charge (SPC) water model and its potential energy includes a 12-6 Lennard-Jones
potential:

\begin{multline}
E_{ClayFF} = K_{r,ij}(r_{ij}-r_{0,ij})^2 + K_{\theta,ij}(\theta_{ij}-\theta_{0,ij})^2 \\ 
+\frac{1}{4\pi\varepsilon_{0}\varepsilon_{r}}\sum_{ij}\frac{q_{i}q_{j}}{r_{ij}} +\sum_{ij}\varepsilon_{0,ij}\left[\left(\frac{\sigma_{ij}}{r_{ij}}\right)^{12}- \left(\frac{\sigma_{ij}}{r_{ij}}\right)^6\right]
\label{eq:potential_clayff}
\end{multline}

For dehydrated species, its interaction potential is the sum of a 12-6 LJ
potential and electrostatic interactions. ClayFF does not account explicitly for
covalent bonding in silicates, and larger charges are used for silicon atoms
(2.1e) when compared to IFF (1.0e). The ionic nature of these bonds is thus
overestimated in ClayFF, and impacts substantially computed surface tension
\cite{Mishra2013}. We also expect an influence of this assumption on the elastic
behaviour \cite{Heinz2013}. Results in better agreement with experimental
measurements are expected with IFF, for which parameters were optimized
especially for \ce{C3S} \cite{Mishra2013}. Every simulation in the present paper
were performed with the LAMMPS simulation code \cite{Plimpton1995}. A
\SI{12}{\angstrom} cutoff, and an Ewald summation with precision of
\num[retain-unity-mantissa = false]{1e-5} were adopted for short-range and
long-range interaction, respectively. 

\section{Mechanical properties}

\subsection{Methods}

Elastic properties of solids are generally computed by applying a strain or a
stress in the desired directions and by determining the strain-stress or
strain-energy relations. Two type of methods are used and discussed in this work: static optimization methods and time integration methods. 

\begin{enumerate}
  \item Static optimization methods are typically applied at \SI{0}{\kelvin}, or
  where anharmonical vibrations can be neglected, although lattice vibration
  frequency can be included through quasi-harmonic approximation techniques
  \cite{Qomi2017}. In this case, a small strain $\Delta \varepsilon_j$ is
  applied positively and negatively in each direction $j$:
  \begin{equation}
  \begin{split}
  C_{ij}^+ = - \frac{\sigma_i(\Delta\varepsilon_j) - \sigma_i(0)}{\Delta\varepsilon_j} \\
  C_{ij}^- = \frac{\sigma_i(-\Delta\varepsilon_j) - \sigma_i(0)}{\Delta\varepsilon_j}
  \end{split}
  \end{equation}
  The stiffness constants can be obtained by averaging $C_{ij}^+$ and
  $C_{ij}^-$ and the symmetric constants:
  \begin{equation}
   C_{ij} = \frac{C_{ij}^+ + C_{ij}^- + C_{ji}^+ + C_{ji}^-}{4} \label{eq:stiffness_constants}
  \end{equation}
  This method performs quick calculation, minimizing the energy of the system
  before and after application of a small strain $\Delta \varepsilon$. However,
  it does not provide the stress-strain behavior nor give a prediction of the
  failure point. This computational scheme can be extended by applying a strain
  on multiple steps followed by an energy minimization after each step. The
  stiffness constants are therefore obtained by linear regression on the desired
  strain range. 
  
  For a system of particles with a volume $V$, the stress components can be
  computed as the sum of the kinetic and virial terms over the $N$ particles:
  \begin{equation}
    \sigma_{ij} = \frac{\sum^N_k m_k v_{ki} v_{kj}}{V} + \frac{\sum^N_k r_{ki} f_{kj}}{V} \label{eq:stress_atom}
  \end{equation}
  where $i$ and $j$ are the directions $x$, $y$ and $z$. $m_k$, $r_{ki}$,
  $v_{ki}$ are the mass, position and velocity respectively, and $f_{kj}$ is the
  force applied on the particle $k$. In the case of a molecular mechanics (MM)
  optimization, the kinetic term is zero. 
  
  \item Time integration methods use equilibrium MD (EMD) or non-equilibrium
  (NEMD) simulations to compute the deformation of the simulation box while
  controlling the stress or vice versa. In EMD, the equilibrium is reached
  before each production run, which provides time-averaged values of the
  computed properties. In NEMD, the strain, or the stress, is changed
  continuously during the run. This is convenient because only a single run is
  needed. However, the strain/stress rate may influence the result. The
  simulation can be either strain or stress controlled. In the first case, a
  strain rate is applied on the desired direction and with a fixed stress
  (usually \SI{0}{\giga\pascal}) on the other directions. In the second case, a
  stress rate is applied in one direction while keeping the others at
  \SI{0}{\giga\pascal}. The simulation box is thus allowed to relax in the other
  directions.
\end{enumerate}

The elastic properties of \ce{M1} and \ce{M3 C3S} were first computed from the
static MM calculation method on unit cells. The enthalpy of the cell was
minimized at \SI{0}{\giga\pascal}, allowing free movement of atoms and cell
parameters. Then a deformation was applied in the desired direction and the
energy of the system was minimized, allowing the atoms to move while fixing the
cell parameters. The process was repeated negatively and positively in each
direction, to calculate the 21 components of the stiffness matrix according to
the \cref{eq:stiffness_constants}. The unit cells experienced maximal
deformations of \num{\pm0.2}, with increments of \num{1e-4}, but the values
$C_{ij}^{\pm}$ were obtained by linear fitting on values from zero to \num{\pm
0.02} deformation. Homogeneous values of bulk and shear moduli for large
crystals randomly dispersed were obtained by calculating Reuss and Voigt bounds.
The Voigt-Reuss-Hill (or VRH) estimation for monoclinic crystals is obtained as
the arithmetic average of Voigt and Reuss bounds on bulk and shear modulus
\cite{Wu2007, Fu2017, Bernard2018}.

In order to determine elastic properties at finite temperature supercells of
1296 atoms (\num{1x4x2} and \num{2x4x3}, with dimensions
\SI{27.87x28.24x24.52}{\angstrom} and \SI{24.47x28.29x27.89}{\angstrom}, for
\ce{M1} and \ce{M3 C3S} respectively) were created from the unit cells presented
in \cref{fig:unitcells}. For each polymorph, three replicas were created by
using different seeds for the initial velocities of atoms. Equilibration runs
were performed during \SI{500}{\ps} at \SI{300}{\kelvin} in the NpT ensemble at
hydrostatic pressure $\sigma$ varying between 0 and \SI{15}{\giga\pascal},
followed by a production run of \SI{1}{\ns}. Nose-Hoover thermostat and barostat
\cite{Nose1984,Hoover1985} were employed with the Verlet algorithm
\cite{Verlet1967} to integrate Newton's equations of motion. Long-range
interaction were computed with an Ewald summation with precision of
\num[retain-unity-mantissa = false]{1e-5} and a cutoff of \SI{10}{\angstrom} was
applied for van der Waals interactions.

The bulk modulus $K$ was calculated from EMD simulations in the NpT ensemble with incremental equilibrium pressure. In NEMD simulations, the supercells underwent \SI{1}{\ns} runs of compression and tension at a strain rate of \SI{1e8}{\per\s} up to \SI{20}{\percent}, while maintaining the pressure to \SI{0}{\giga\pascal} in the other directions. The stress components were computed from \cref{eq:stress_atom}. No noticeable influence was reported for rates of one order of magnitude above and bellow \SI{1e8}{\per\s}. This is predictable because the resulting dislocation velocity is \SI{\sim 0.28}{\meter\per\second} for the largest dimension. This dislocation velocity is large when compared to macroscale tests, but is negligible compared to the velocity of acoustic waves in \ce{C3S}. Based on the values of bulk modulus $K$, Poisson's ratio $\nu$ and density $\rho$ from previous acoustic measurements, compressive and shear waves are calculated as \num{7200} and \SI{3700}{\meter\per\second} respectively \cite{Boumiz1997}. This ensures that atoms will respond instantaneously to the deformation of the simulation box \cite{Elder2016,Teich-McGoldrick2012}. Elastic parameters were calculated by the direct relations, where $i \neq j$ are the $x$, $y$ and $z$ coordinates:
\begin{align}
E_{ii} &= \frac{\sigma_{ii}}{\varepsilon_{ii}} \notag \\ 
G_{ij} &= \frac{\sigma_{ij}}{\varepsilon_{ij}} \\
\nu_{ij} &= - \frac{\varepsilon_{ii}}{\varepsilon_{jj}} \notag
\end{align}

\subsection{Results and discussion}

\begin{figure}[!htbp]
  \centering
  \includegraphics[width=0.8\textwidth]{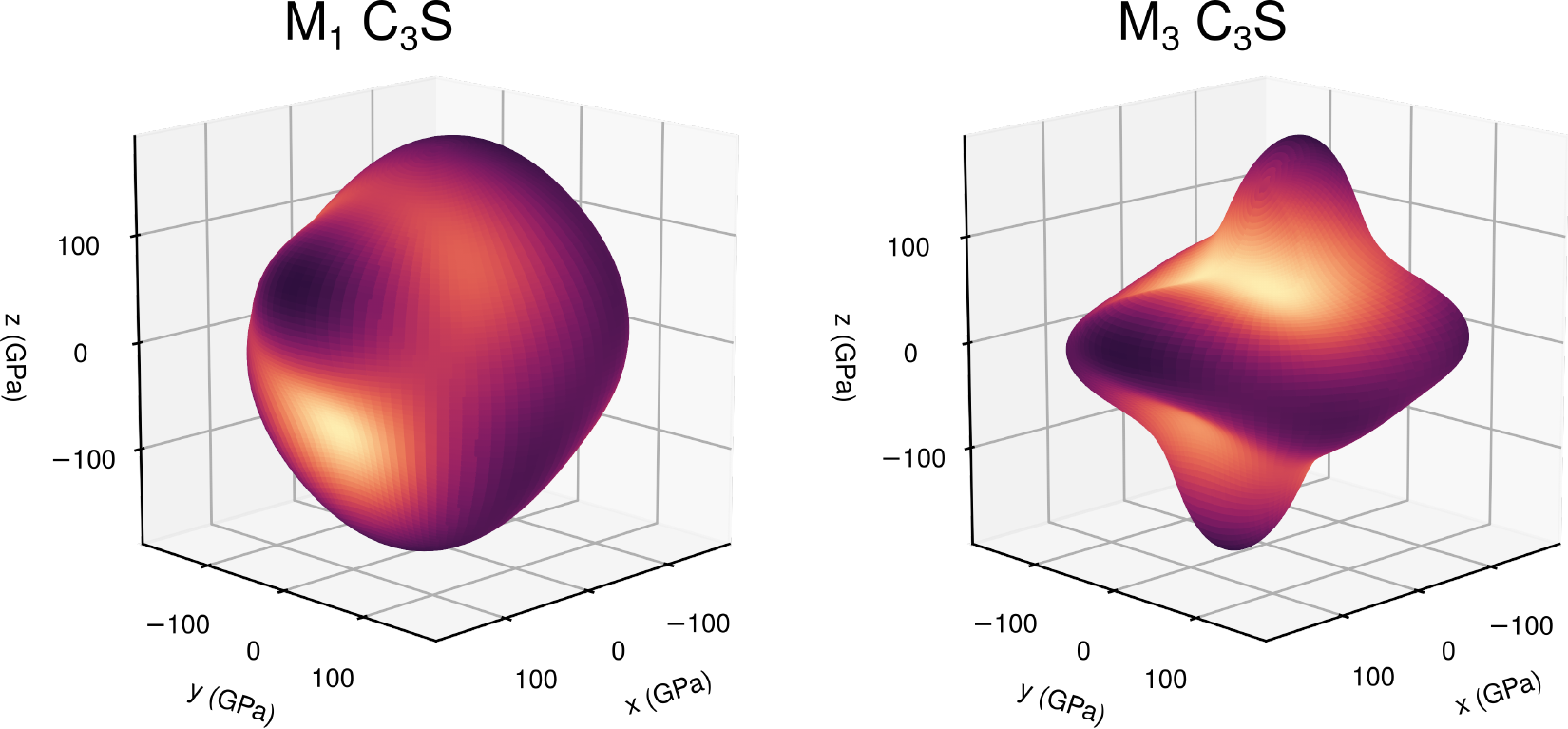}
  \caption{Spacial distribution of Young's modulus for the \ce{M1 C3S}, and \ce{M3 C3S} computed with IFF using the static MM method.}
  \label{fig:elastic_surface}
\end{figure}

The stiffness constants, homogenized stiffness constants and elastic moduli of
\ce{M1} and \ce{M3 C3S}, computed by static MM method, as well as experimental results from literature, are reported in \cref{tab:elastic}.
\begin{table}[!htbp]
  \centering
  \sisetup{separate-uncertainty = false}
  \centering
    \begin{tabular}{*7c}
    \toprule
        & \multicolumn{2}{c}{\ce{M1}} & \multicolumn{2}{c}{\ce{M3}} & \multicolumn{2}{c}{Exp.}  \\
    \cmidrule{2-5}  
        & IFF & ClayFF & IFF & ClayFF & Boumiz et al.$^a$ \cite{Boumiz1997} & Velez et al. \cite{Velez2001} \\
    \cmidrule{2-5}         
    $C_{11}$ & \num{185.2(2)}   & \num{89.3(2)}  & \num{219.6(3)}  & \num{118.6(2)} & & \\
    $C_{12}$ & \num{62.11(9)}   & \num{29.6(4)}  & \num{77.54(9)}  & \num{34.81(7)} & & \\
    $C_{13}$ & \num{61.73(8)}   & \num{27.83(7)} & \num{52.79(7)}  & \num{35.98(5)} & & \\ 
    $C_{15}$ &\num{17.35(3)}   & \num{-0.12(7)} & \num{4.0(1)}    & \num{17.69(4)} & & \\
    $C_{22}$ & \num{216.1(3)}   & \num{112.6(5)} & \num{216.0(3)}  & \num{85.9(3)} & &  \\
    $C_{23}$ & \num{70.77(9)}   & \num{33.31(8)} & \num{52.72(8)}  & \num{27.95(8)} & & \\
    $C_{25}$ & \num{-8.38(2)}   & \num{-4.7(3)}  & \num{-21.7(9)}  & \num{1.39(4)} & & \\
    $C_{33}$ & \num{212.7(3)}   & \num{100.2(3)} & \num{189.6(2)}  & \num{95.4(1)} & & \\
    $C_{35}$ & \num{-9.2(3)}    & \num{-7.93(3)} & \num{-34.04(2)} & \num{6.65(3)} & & \\ 
    $C_{44}$ & \num{66.54(3)}   & \num{34.93(4)} & \num{37.0(7)}   & \num{33.01(5)} & & \\
    $C_{46}$ & \num{-4.954(7)}  & \num{-3.958(8)} & \num{-6.39(8)}  & \num{3.28(5)} & & \\
    $C_{55}$ & \num{66.44(3)}   & \num{31.35(4)} &  \num{43.4(6)}   & \num{38.22(4)} & & \\
    $C_{66}$ & \num{65.17(3)}   & \num{32.21(4)} &  \num{67.65(7)}  & \num{32.16(6)} & & \\
    \cmidrule{2-5} 
    $K$      & 111.1    &   53.2  & 104.7  &  53.9 & \num{105.2(5)} &  \\  
    $G$      & 66.7     &   33.5  & 54.3   &  33.5 & \num{44.8(6)} &  \\
    $E$      & 166.8    &   83.0  & 139.0  &  83.2 & \num{117.6(8)} & \num{147(5)}$^b$/\num{135(7)}$^c$ \\
    $\nu$     & 0.250    &   0.240 & 0.279  &  0.243 & \num{0.314(17)}  &  \\ 
    $E_{11}$  & 151.5    &   77.2  & 176.9  &  91.2 & &  \\
    $E_{22}$  & 180.6    &   96.2  & 174.6  &  71.7 & &  \\
    $E_{33}$  & 177.3    &   84.4  & 147.2  &  80.4 & &  \\
    $E_{44}$   & 66.2     &   34.4  &  36.3   &  32.7 & &  \\
    $E_{55}$  & 63.1     &   30.6  & 34.9   &  35.3 & &  \\
    $E_{66}$  & 64.8     &   31.7  & 66.5   &  31.8 & &  \\
    \bottomrule
    \end{tabular}
    \caption{Stiffness constants and elastic moduli of \ce{M3 C3S} obtained by static MM, in \si{\giga\pascal}. $^a$ Acoustic measurements. $^b$Resonance frequency. $^c$ Nanoindentation.}
    \label{tab:elastic}
    \sisetup{separate-uncertainty}
\end{table}
ClayFF tends to underestimate by a factor of approximately 2 the stiffness
constants and thus the elastic moduli. This very probably results from the
non-bonded nature of atomic interactions in ClayFF, where the covalent nature of
O-Si bonds in silicates is underestimated, thus decreasing their stiffness.
Previous calculations on the same \ce{M3} unit cell, via second derivative of
the binding energy with the GULP code, lead to very similar results
\cite{Shahsavari2016}. The homogenized elastic moduli computed with IFF agree
relatively with experiments. The values obtained for \ce{M3 C3S} are smaller
than for \ce{M1} and close to recent results from DFT calculations on the \ce{T1
C3S} \cite{Laanaiya2019}. Very similar results were obtained by Manzano et al.
\cite{Manzano2009}, employing the Buckingham FF and a \ce{M3 C3S} model proposed
by de la Torre et al. \cite{Torre2002}. The lowest elastic modulus is obtained
in the $x$ and $z$ direction for the \ce{M1} and \ce{M3} polymorph respectively.
This result is predictable because of the correspondence of the $c$ parameter of
the \ce{M1} unit cell with the a parameter of the \ce{M3} unit cell. Spacial
distributions of Young's modulus were plotted with the ELATE open-source Python
package \cite{Gaillac_2016} and reveal a much more anisotropic elasticity for \ce{M3} than for \ce{M1 C3S} (see \cref{fig:elastic_surface}). 

EMD calculations where performed with the IFF at 0, 1, 2.5 and \SI{5}{\giga\pascal}. The bulk moduli obtained by linear fitting of the
hydrostatic pressure with respect to the volume variation are
\SI{101(2)}{\giga\pascal} and \SI{103(2)}{\giga\pascal} for \ce{M1} and \ce{M3
C3S} polymorphs, respectively. A larger difference with static MM calculation is
observed for \ce{M1}. This could rely on the fact that the \ce{M1} model, which
is not averaged, experienced a structural relaxation during the MD run.

\begin{figure}[!htbp]
  \centering
  \includegraphics[width=0.6\textwidth]{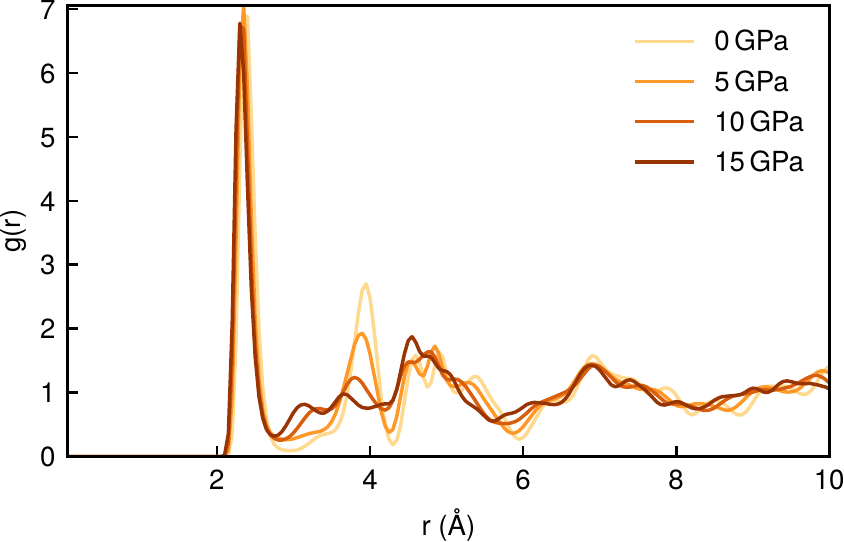}
  \caption{Radial distribution function of Ca-Os pairs as a function of the hydrostatic pressure for \ce{M3 C3S}, with IFF. The RDF obtained for the \ce{M1} polymorph is very similar.}
  \label{fig:rdf_iso}
\end{figure}

As an important feature of the \ce{C3S}, the change in coordination between calcium cations Ca and oxygen in silicates (Os) as a function of the hydrostatic pressure was analyzed by radial distribution function (RDF) (see \cref{fig:rdf_iso}).
The increasing hydrostatic pressure seems to influence the \ce{C3S} structure at short range (\SI{\sim 4}{\angstrom}). The second coordination shell is flattened and shifted to the left by the effect of the pressure. The same behavior is observed for both polymorphs.

\begin{figure}[!htbp]
  \centering
  \includegraphics[width=\textwidth]{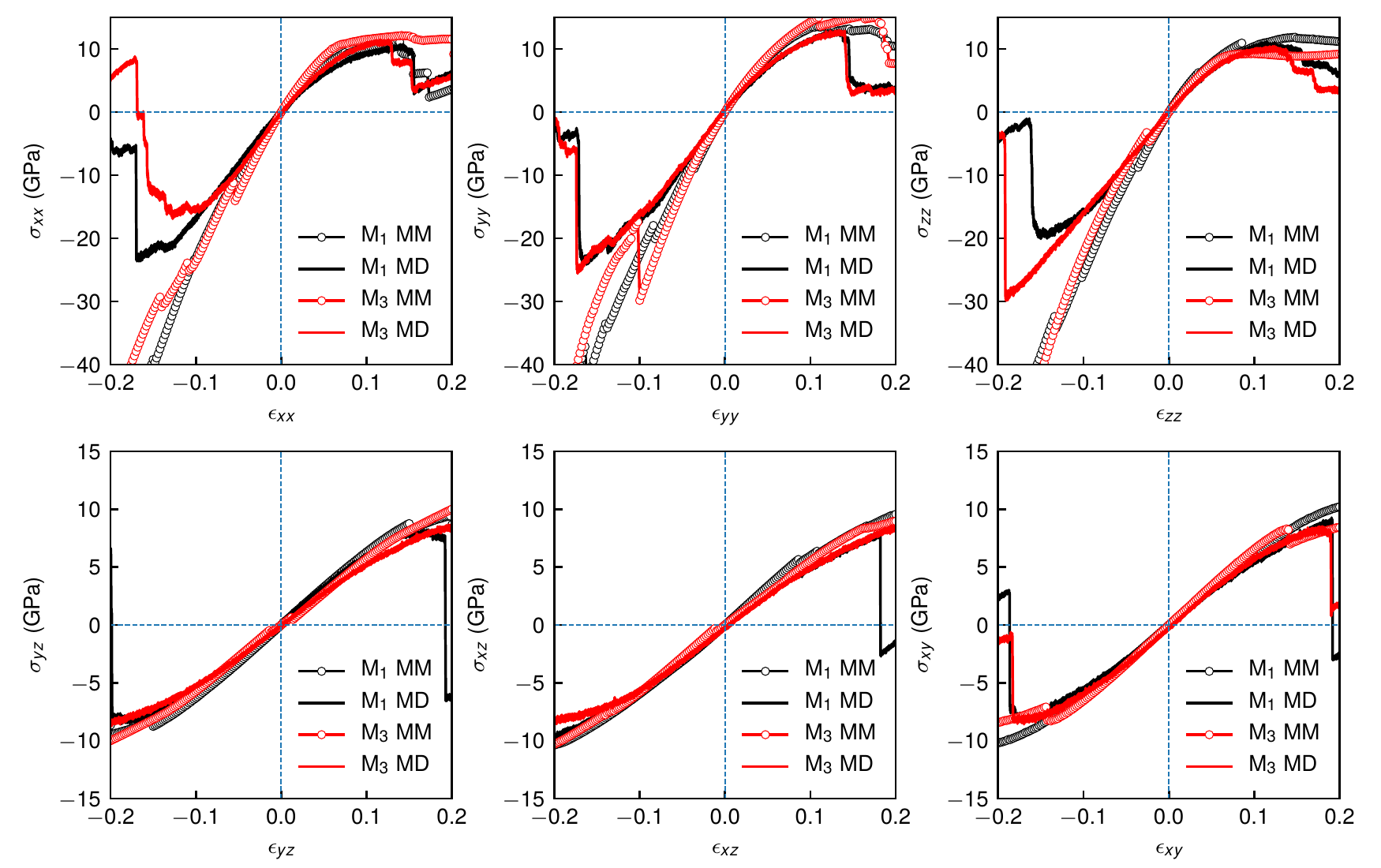}
  \caption{Stress-strain curves obtained by NEMD and MM calculations with IFF.}
  \label{fig:strain_nemd}
\end{figure}

The stress-strain curves for NEMD and static MM simulations are plotted in
\cref{fig:strain_nemd}. The general behavior of the \ce{M1} and \ce{M3}
polymorphs seems similar. However, the compressive strength seems to be larger
for the \ce{M1} polymorph in the $x$ direction, and for the \ce{M3} polymorph in
the $z$ direction. The structural correspondence of the (001) direction for the
\ce{M1} unit cell with the (100) direction for the \ce{M3} unit cell explains
this result.As already noticed for the elastic behaviour (see
\cref{fig:elastic_surface}), a more isotropic yield behaviour is observed for
\ce{M1} when compared to \ce{M3}. The difference in compressive strength along
the $x$ and $z$ directions is greater for \ce{M3 C3S} ($\sim$ 15 and
\SI{30}{\giga\pascal}) than for \ce{M1 C3S} ($\sim$ 21 and
\SI{16}{\giga\pascal}). The elastic moduli obtained from these NEMD simulation
are presented in \cref{tab:el_nemd}. They are in good agreement with values from
previous stress controlled NEMD simulations on \ce{M3 C3S} \cite{Mishra2013} and
\ce{T1 C3S} \cite{Tavakoli2016}.
\begin{table}[!htbp]
  \sisetup{separate-uncertainty = false}
  \centering
  \begin{tabular}{*4c}
  \toprule
              & \ce{M1} & \ce{M3} & \ce{M3}$^a$ \cite{Mishra2013} \\
  \midrule
  $E_{11} = E_{xx}$    & \num{141.8(2)}  & \num{168.5(2)}    & \num{152(6)}  \\ 
  $E_{22} = E_{yy}$    & \num{164.4(2)}  & \num{166.0(2)}    & \num{176(3)}  \\
  $E_{33} = E_{zz}$    & \num{155.8(2)}  & \num{154.9(1)}    & \num{103(11)} \\
  $E_{44} = G_{yz}$    & \num{62.6(5)}   & \num{57.16(6)}    & \\
  $E_{55} = G_{xz}$    & \num{59.22(7)}  & \num{57.42(6)}    & \\
  $E_{66} = G_{xy}$    & \num{59.77(5)}  & \num{64.35(6)}    & \\
  $\nu_{12} =\nu_{xy} $  & \num{0.220(1)}  & \num{0.297(1)} & \num{0.303(42)}
  \\ 
  $\nu_{13} =\nu_{xz} $  & \num{0.236(2)}  & \num{0.227(2)}  & \num{0.273(43)}
  \\ 
  $\nu_{21} =\nu_{yx} $  & \num{0.241(2)}  & \num{0.290(1)} & \num{0.225(21)}
  \\ 
  $\nu_{23} =\nu_{yz} $  & \num{0.239(2)}  & \num{0.215(1)}  & \num{0.197(27)}
  \\ 
  $\nu_{31} =\nu_{zx} $  & \num{0.245(2)}  & \num{0.200(1)}    & \num{0.372(41)}
  \\ 
  $\nu_{32} =\nu_{zy} $  & \num{0.230(1)}  & \num{0.200(1)}    & \num{0.299(58)}
  \\ 
  \bottomrule
  \end{tabular}
  \caption{Elastic moduli obtained from NEMD simulations with IFF. $^a$Results from previous stress controlled NEMD.}
  \label{tab:el_nemd}
  \sisetup{separate-uncertainty}
\end{table}

MM simulations results in a stiffer elastic behaviour than for MD simulations,
due to the larger structural relaxation induced by the thermal motion.
Nonetheless, the values obtained by these two methods are in good agreement. The
static calculation method provides a good representation of elastic properties
and is very fast, but does not allow to assess the yield stress properly, in
particular in compression, since no relaxation is permitted in the
transversal directions. This restriction causes hardening for negative strains.
The yield stress could be assessed by enthalpy minimization, but such
calculation is not trivial and the calculation can easily stuck in a local
minima because the objective function is changing while the simulation box
dimensions change.

\section{Thermal properties}

During its lifetime, concrete undergoes temperature changes. The thermal
expansion and contraction of concrete as temperature increases and decreases, is
influenced by the aggregate type, the cement type, and the water/cement ratio.
Although the aggregate type has the larger influence on the expansion and
contraction of concrete, the thermal properties of hydrated and dry cement is of
great interest. Thermal cracking of concrete generally occurs during the first
days after casting. During the exothermic hydration of cement, the temperature
rises, and drops faster on the surface than in the bulk. The surface tends to
contract with the cooling and stress arises because the bulk remains hot,
resulting in cracks. Naturally, this phenomenon occurs more likely in larger
volumes. The assessment of vibrational spectra and specific heat of cement phase
is a first step toward a microstructural modelling and further understanding of
heat propagation in the cement paste. 

\subsection{Methods}

In the canonical ensemble, one can derive the fluctuation relationship between
specific heat and internal energy $E$:
\begin{equation}
C_V = \frac{1}{kT^2} \expval{(\Delta E)^2}
\end{equation}
where $k$ is the Boltzmann constant. By using fluctuation methods, the specific
heat can be computed at any temperature with a single, long enough run. However,
these methods rely strongly on the temperature relaxation parameter used to
thermostat the system (and in the case of the NpT ensemble, on the pressure
relaxation parameter) \cite{Hickman2016}. Moreover values obtained by
fluctuation method depends on the time interval used for block averages
\cite{Hickman2016}, often leads to large uncertainties and to bad agreements
with experimental results \cite{Wang2010a}. For this reason, non-fluctuation, or
direct method, is preferred. It consists on running several simulations at
finite temperature and calculating the time average energies for each one. The
specific heat is calculated by definition, as the slope of the internal energy with
respect to the temperature, at the desired temperature. The chosen temperature
increment must be large enough to compute accurately the variation of energy
between each simulation, but small enough, for the fitting to be representative.

The specific heat capacity can also be computed from the velocity
autocorrelation function (VACF) of atoms. Considering a solid made by
quantum harmonic oscillator, the phonon density of states $g(\omega)$ is
proportional to the Fourier transform of the velocity autocorrelation function
of the atoms:
\begin{equation}
g(\omega) = \frac{1}{3NkT} \int_{-\infty}^{+\infty} \sum^N_{i=1} \expval{\vb{v}_i(t) \vdot \vb{v}_i(0)} e^{i \omega t} \dd t
\label{eq:dos}
\end{equation}
where k is the Boltzmann constant, $N$ is the number of atoms and $T$ is the temperature of the system. The occupational states of phonons follows a Bose-Einstein distribution $f_{BE}$ and at energy largely below the Debye temperature, the internal energy of the system can be reduced to the vibrational energy $E_v$ \cite{Qomi2015,Atkins2009}:
\begin{equation}
E_v = \int_0^{+\infty} \hbar \omega \qty(g(\omega) f_{BE}(\omega) + \frac{1}{2}) \dd \omega
\end{equation}
The specific heat $c_v$ in $3Nk$ units is calculated as the partial derivative of internal energy with respect to temperature:
\begin{equation}
c_v = \frac{\int_0^{+\infty} \cfrac{u^2 e^u}{(1-e^u)^2} g(\omega) \dd \omega}{\int_0^{+\infty} g(\omega) \dd \omega}
\end{equation}
where $u = \hbar \omega / kT$. One has to note that the phonon spectrum computed this way is semi-classical and is different from the quantum phonon spectrum \cite{Winkler1992}. Although the \cref{eq:dos} includes quantum effects, the density of states is obtained by classical MD. We consider that this method is applicable at standard temperature and that anharmonic interactions are negligible. 

Most of the time, experimental measurement of specific heat capacities is
performed at constant pressure. From thermodynamics, the specific heat
capacities at constant volume and pressure $c_v$ and $c_p$, are related by the
equation:
\begin{equation}
c_p - c_v = T \frac{\alpha ^2}{\rho \beta} \label{eq:cp-cv}
\end{equation}
where $\rho$ is the density, $\alpha = (1/V)(\pdv*{V}{T})_p$ is the thermal
expansion coefficient, and $\beta = -(1/V)(\pdv*{V}{P})_T$ is the
compressibility, inverse of the bulk modulus $K$.

Specific heat and thermal expansion coefficient were computed on three replicas
of \ce{C3S} supercells. The simulations were performed in the NpT ensemble with
the same MD parameters as previously. The systems were relaxed during
\SI{0.5}{\ns}, and the data were collected for \SI{1}{\ns}. Within the direct
method, the specific heat $c_p$ was computed by linear fitting of the enthalpy
with respect to the temperature at five points around the temperature of
interest (e.g., 280, 290, 300, 310 and \SI{320}{\kelvin} to compute the specific
heat at \SI{300}{\kelvin}). The same method was employed to calculate the
expansion coefficient, fitting the volume variation with respect to the
temperature. To avoid the external influence of thermostating or barostating,
the relaxed systems were equilibrated for \SI{500}{\ps} in the NVE ensemble,
before running simulation of \SI{100}{\ps}, dumping the trajectory at each time
step to be able to observe high vibrational frequencies. For the calculation
using the ClayFF, a geometric mixing rule for LJ parameters was used in place of
the original arithmetic mixing. Indeed, this mixing rule provides more accurate
value of density obtained during NpT simulations. The VACF were computed on ten
correlation windows of \SI{10}{\ps} on three replica for each polymorph.

\subsection{Results and discussion}

\begin{figure}[!htbp]
  \centering
  \includegraphics[width=0.8\textwidth]{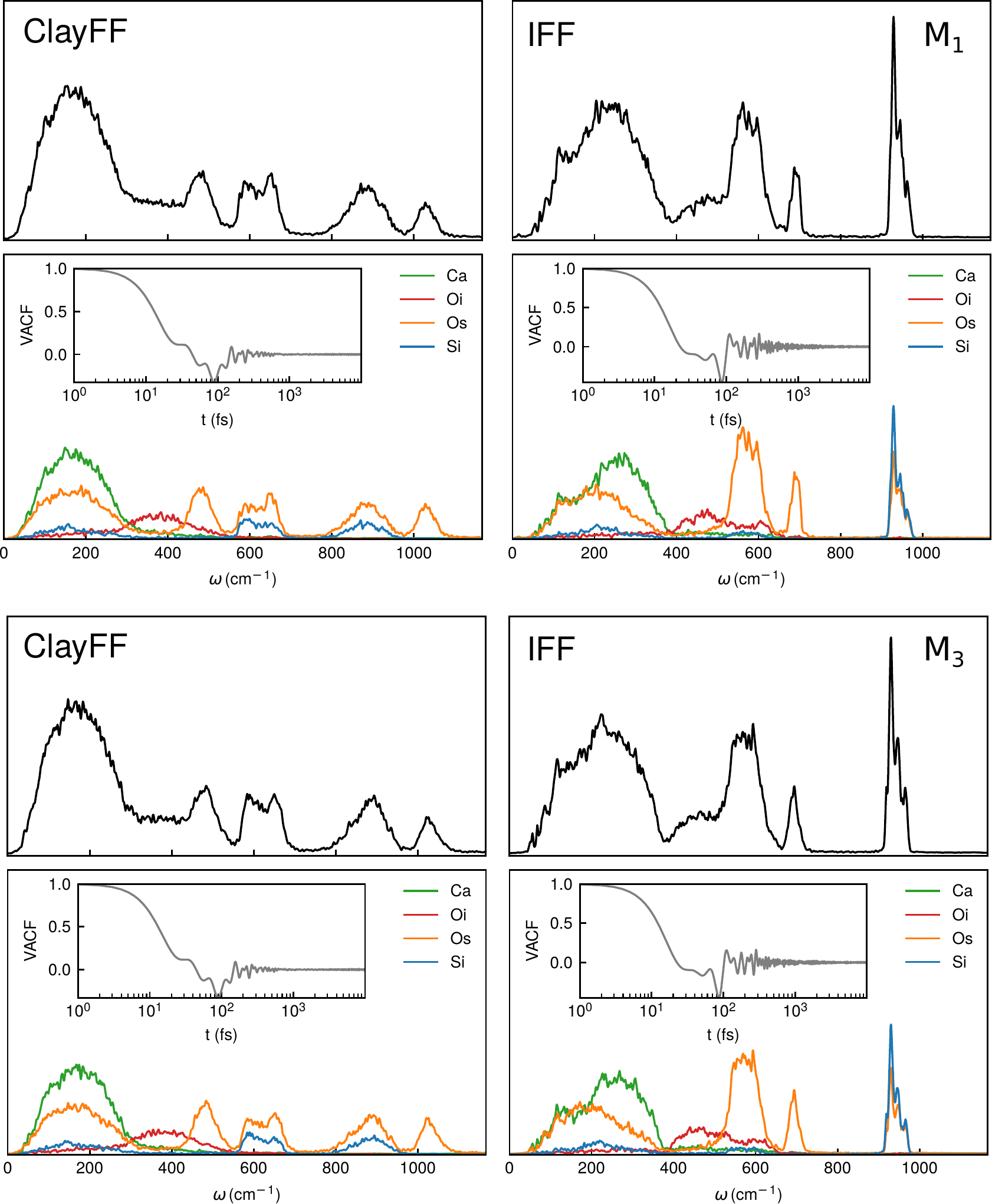}
  \caption{Total phonon density of states (top), and partial density of states with VACF in insets (bottom) of \ce{M1} and \ce{M3 C3S} obtained with IFF and ClayFF.}
  \label{fig:vdos}
\end{figure}
\begin{table*}[!hbt]
  \centering
\sisetup{separate-uncertainty = false}
  \begin{tabular}{*6c}
  \toprule
        & Temperature (\si{\kelvin}) & $\alpha$ (\si{\per\kelvin}) & $\beta$ (\si{\per\pascal}) & $\rho$ (\si{\gram\per\centi\meter\cubed}) & $c_p-c_v$ (\si{\joule\per\gram\kelvin}) \\
  \midrule
  \ce{M1} & 200 & \num{4.6(18)e-5} & \num{9.9(2)e-12} & \num{3.147(5)} & \num{0.013(14)}  \\
          & 300 & \num{4.4(15)e-5} & \num{9.7(3)e-12} & \num{3.160(4)} & \num{0.019(13)}  \\
          & 400 & \num{5.0(23)e-5} & \num{10.0(2)e-12} & \num{3.147(6)} & \num{0.032(30)}  \\
  \midrule  
  \ce{M3} & 200 & \num{3.6(14)e-5} & \num{9.7(2)e-12} & \num{3.139(5)} & \num{0.009(8)} \\
          & 300  & \num{4.0(16)e-5} & \num{9.7(3)e-12} & \num{3.151(4)} & \num{0.016(13)} \\
          & 400  & \num{4.8(15)e-5} & \num{9.5(2)e-12} & \num{3.128(6)} & \num{0.031(21)} \\
  \bottomrule
  \end{tabular}
\sisetup{separate-uncertainty}
\caption{Thermal properties of \ce{M1} and \ce{M3 C3S}.}
\label{tab:npt_thermo}
\end{table*}

The phonon density of states (DOS) obtained from simulations in the NVE ensemble
are presented in \cref{fig:vdos}.The phonon DOS obtained for \ce{M1} and \ce{M3}
are almost identical, so as the resulting specific heat. Thus the structural
difference between both polymorph does not influence their thermal properties.
However, the force field does affect the results. The main difference between
the phonon DOS obtained with IFF and ClayFF relies principally in their
description of bonds in silicate. In the case of IFF, where Si-O bonds are
described by harmonic oscillators in addition to the short and long range
description, the partial DOS (PDOS) of Os and Si atoms form a sharp peak near
\SI{935}{\per\centi\meter}, in agreement with infrared spectroscopy measurements
\cite{Hughes1995}, while the in-plane bending vibration of Os-Si-Os angle is
measured as a band bellow \SI{500}{\per\centi\meter} \cite{Hughes1995}. This
bending contribution happens at larger frequencies in our results (near
\SI{550}{\per\centi\meter}). Wave numbers $\omega$ bellow
\SI{500}{\per\centi\meter} are associated to stretching between calcium and
oxygen atoms \cite{Yu2004,Qomi2015}, in agreement with the PDOS obtained with
both IFF and ClayFF. For ClayFF, the purely non-bonded description of Si-Os
bonds allows for more degrees of freedom of atoms. The PDOS obtained for Si and
Os atoms are thus more sparse. Generally, for the IFF, a shift of the DOS is
observed towards higher vibrational frequencies. No significant variation of the
DOS was observed between 200, 300 and \SI{400}{\kelvin}. The error on the
calculation of the isobaric specific heat $c_p$ are mostly related to the
$c_p-c_v$ quantity, calculated from simulations in the NpT ensemble. The values
of $c_p-c_v$ were calculated at 200, 300 and \SI{400}{\kelvin} and extrapolated
linearly, because this quantity vary proportionally with temperature (see
\cref{eq:cp-cv}). The thermal expansion coefficient $\alpha$, the compressibility
$\beta$, and the density $\rho$, computed from simulations in the NpT ensemble
are presented in \cref{tab:npt_thermo}. 

\begin{figure}[!htbp]
  \centering
  \includegraphics[width=0.6\textwidth]{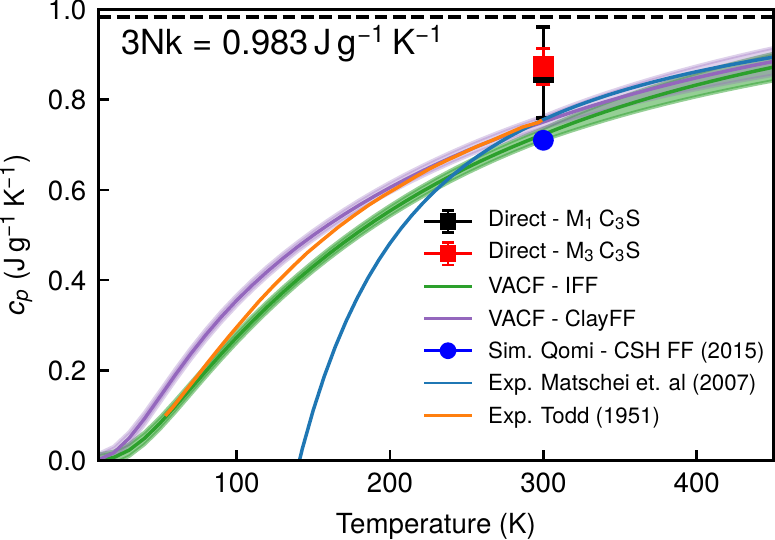}
  \caption{Specific heat of \ce{M1} and \ce{M3 C3S} obtained by the direct and VACF method, plotted as function of the temperature. The transparent areas represent the error. Previously computed value from VACF calculation, as well as fitting of experimental measurements \cite{Matschei2007} and direct measurements \cite{Todd1951} are plotted in addition to the results.}
  \label{fig:specific_heat}
\end{figure}

\sloppy
The specific heat $c_p$ obtained for \ce{M1} and \ce{M3 C3S} are plotted with
respect to the temperature in \cref{fig:specific_heat}. The values obtained at
\SI{300}{\kelvin} by the direct method are \num{0.86(10)} and
\SI{0.87(4)}{\joule\per\gram\per\kelvin} for \ce{M1} and \ce{M3 C3S}
respectively, which is much larger than experimental measurements. As for the
phonon DOS, no significant variation of $c_p$ was found between \ce{M1} and
\ce{M3}. The VACF method with ClayFF results in $c_p =
\SI{0.751(13)}{\joule\per\gram\per\kelvin}$, which is very close to experimental
values of \SI{0.756}{\joule\per\gram\per\kelvin} \cite{Matschei2007} and
\SI{0.753}{\joule\per\gram\per\kelvin} \cite{Todd1951}. The IFF provided a value
slightly lower than experimental measurements:
\SI{0.723(13)}{\joule\per\gram\per\kelvin}.

\section{Cleavage energies and equilibrium shapes}

In the current section, the cleavage energies of \ce{M1} and \ce{M3 C3S} were calculated from energy difference of cleaved and unified slabs. From these energies, the crystal shapes of both polymorphs were constructed using the Wulff construction method \cite{Wulff1901}. 

\subsection{Methods}

\begin{table*}[!htb]
  \centering
  \begin{tabular}[t]{cc}
    \multicolumn{2}{c}{\ce{M1 C3S}} \\
    \toprule
    {Miller index} & Cleavage energy (\si{\joule\per\meter\squared}) \\
    \midrule
    (100)         & \num{1.04(4)} \\ 
    (010)         & \num{1.41(4)} \\
    (040)         & \num{1.76(3)} \\
    (003)         & \num{1.20(3)} \\
    (008)         & \num{1.20(4)} \\
    (110)         & \num{1.45(4)} \\
    (101)         & \num{1.51(3)} \\
    (011)         & \num{1.17(4)} \\
    (111)         & \num{1.43(3)} \\
    \bottomrule
  \end{tabular}
  \hspace{0.5cm}
  \begin{tabular}[t]{cc}
    \multicolumn{2}{c}{\ce{M3 C3S}} \\
    \toprule
    {Miller index} & Cleavage energy (\si{\joule\per\meter\squared}) \\
    \midrule
    (100)         & \num{1.39(3)} \\ 
    (300)         & \num{1.14(3)} \\
    (800)         & \num{1.17(3)} \\
    (010)         & \num{1.55(4)} \\
    (040)         & \num{1.31(4)} \\
    (001)         & \num{1.38(4)} \\
    (002)         & \num{1.45(4)} \\
    (003)         & \num{1.31(4)} \\
    ($00\bar{3}$) & \num{1.33(4)} \\
    (008)         & \num{1.22(4)} \\
    ($00\bar{8}$) & \num{1.22(4)} \\
    \bottomrule    
  \end{tabular}
\caption{Cleavage energies of \ce{M1} and \ce{M3 C3S}. Only the lowest energy plane is given in the (100) direction for \ce{M1}. Results for \ce{M3 C3S} are from our previous study \cite{Claverie2019}.}
\label{tab:cleav_m3}     
\end{table*}

The calculation of cleavage energies for multiple planes of the two monoclinic \ce{C3S} models under study involved creation of cleaved and unified systems. For non-symmetric planes, reorganization of surface ions was performed to minimize superficial dipole moments. For each plane, five unified and cleaved systems were constructed with random distribution of surface species. A \SI{10}{\nm} vacuum was employed in cleaved systems. Series of steep temperature gradients were applied on ions within the uppermost and lowermost atomic layer of slabs in unified and cleaved systems. This method was previously employed and allows to relax the surfaces to the configuration of lower energy \cite{Fu2010,Mishra2013}. The systems with lower energies were selected to performed the calculation over \SI{300}{\ps} runs, after \SI{200}{\ps} equilibration runs. For more details on the method, we refer the reader to ref. \cite{Claverie2019} 

\subsection{Results and discussion}

\begin{figure}[!tb]
  \centering
  \includegraphics[width=0.8\textwidth]{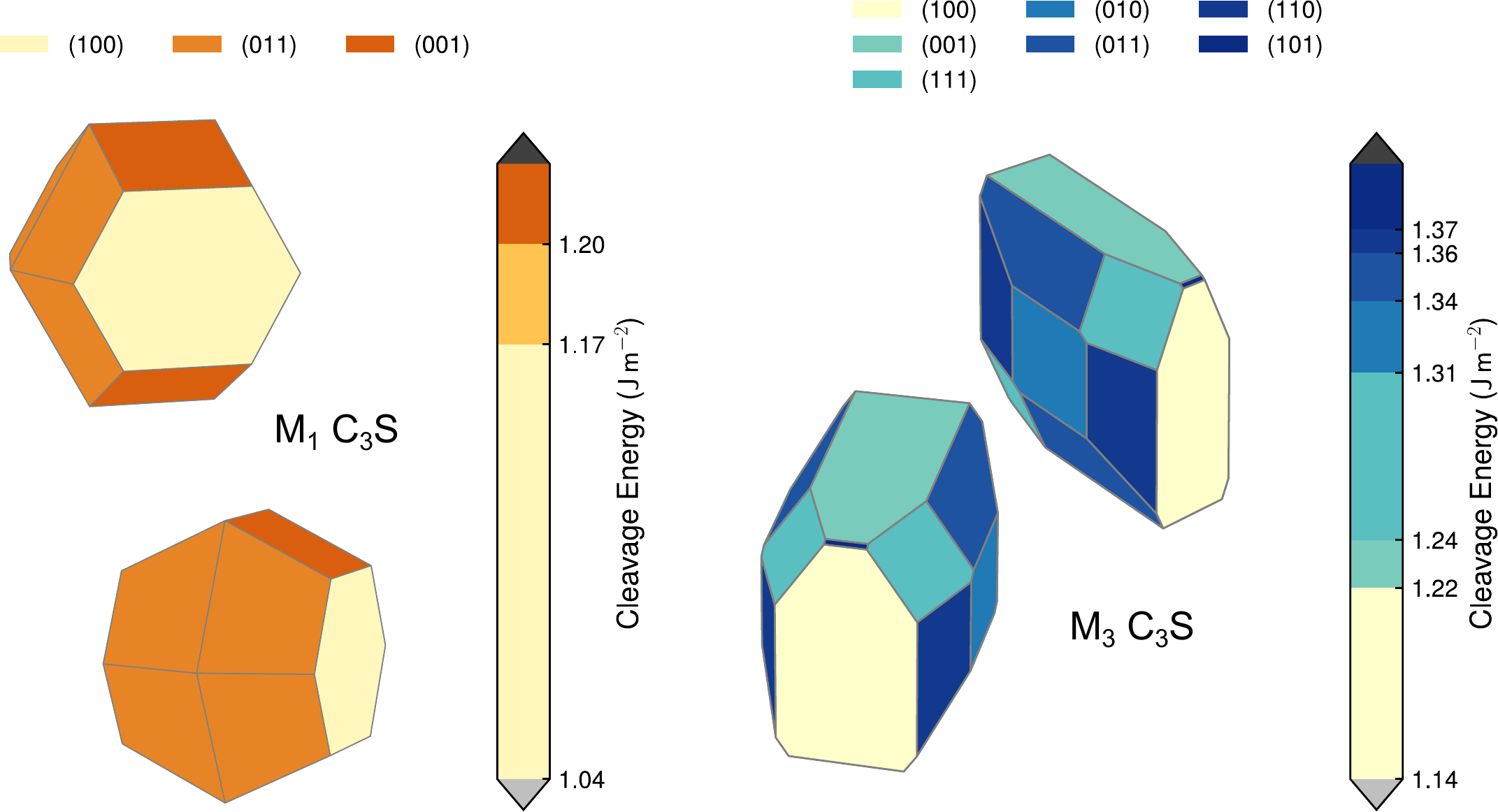}
  \caption{Equilibrium shapes of \ce{M1} and \ce{M3 C3S}}
  \label{fig:shapes}
\end{figure}
  
The cleavage energy was computed classically from \cref{eq:cleavage_energy}
and results are given in \cref{tab:cleav_m3}.

  \begin{equation}
  E_{cleav} = \frac{E_{cleaved} - E_{unified}}{2A} \label{eq:cleavage_energy}
  \end{equation}

The computed values are in the range of 1.14 to
\SI{1.55}{\joule\per\meter\squared} (\SI{1.32}{\joule\per\meter\squared} in
average) for \ce{M3 C3S}, and 1.04 to \SI{1.76}{\joule\per\meter\squared}
(\SI{1.35}{\joule\per\meter\squared} in average) for \ce{M1 C3S}, in good
agreement with previous atomistic simulation studies
\cite{Durgun2014,Manzano2015}. In general, the average values are very close
between the two models. The energies obtained for particular planes vary as a
function of the structure of each polymorph, and are particularly influenced by
coordination between calcium cations and oxygen atoms in silicates. The (100)
direction of the \ce{M1} polymorph corresponds to the (001) direction of the
\ce{M3} polymorph, and vice-versa. For \ce{M1 C3S}, the present calculation
indicates that the lowest energy plane is in the (100) direction, at
\SI{2}{\angstrom} from the origin. The (100) direction of the \ce{M1} polymorph
has many possible cleavage planes and the plane for which the lowest energy was
computed has no equivalent in the \ce{M3} polymorph.

The Wulff construction can give theoretical insights on the shape of a crystal
at equilibrium \cite{Wulff1901}. It is based on the assumption that a crystal
growth in a such way that the Gibbs free energy of its surface is minimized
\cite{Gibbs1928}. This results in a proportional relationship between the
surface energy of a facet and its distance from the crystal center. From the
lowest energy obtained in each direction, the equilibrium shapes of \ce{M1} and
\ce{M3 C3S} in \cref{fig:shapes} were created with the construction algorithm
implemented in the pymatgen library \cite{Tran2016,Ong2013}. In the \ce{M1}
polymorph, the crystal grows only along three planes, while in the \ce{M3},
seven planes are available. Maki related that the equilibrium form of alite is
usually made up of three special forms: one pedion and two rhombohedra
\cite{Maki1986}. The author proposed a morphology similar to the \ce{M1}
obtained by Wulff construction, though more flat and with only one rhombohedra
form. Maki explains that the crystal form of alite changes during
recrystallization from platelet to massive granules with well developed
pyramidal faces (10$\bar{1}$1) and (1$\bar{1}$02) \cite{Maki1983}. The ratio
between the width and the length of the platelet is function of the environment
during the growth. One should note that the obtained equilibrium shapes are not
fully definitive, since they were determined on a relatively finite number of
planes. More calculations would probably refine these shapes, and produce a more
accurate prediction.

\section{Conclusion}

This research aimed to provide knowledge at the atomic scale on the influence of
alite polymorphism on its mechanical and thermal properties as well as on its
equilibrium shape. The two main polymorphs of \ce{C3S} in industrial OPC,
\ce{M1} and \ce{M3}, were investigated. This knowledge may contribute to the
understanding of the influence of alite polymorphism on the variation in
strength of Portland cement. This work also provide input data that are
necessary in microscale modelling of Portland cement hydration and the
development of its mechanical and thermal behavior \cite{Kamali_Bernard_2009,
Bernard2010}. Cleavage energy values may improve our understanding of alite
reactivity and dissolution kinetic, and crystal shapes could help for
identification of polymorphs by SEM. In addition, this study explored and
discussed different calculation methods and compared the performance of two force fields widely used to described cementitious systems.

The elastic constants were calculated by static and dynamical methods. The
moduli found by Voigt-Reuss-Hill homogenization of the stiffness constants
obtained in static calculation with IFF were found in good agreement with
experimental measurements. However, the results obtained with ClayFF
overestimated and underestimated experimental results, respectively. An
isotropic distribution of elastic moduli in space was observed for the \ce{M1}
polymorph, whereas an anisotropic distribution was found for \ce{M3}. The
assessment of stress-strain curves for both polymorphs also indicates a more
anisotropic behaviour for \ce{M3 C3S} regarding the yield compressive stress.
Bulk modulus was obtained by EMD, and elastic, shear modulus, as well as
Poisson's ratio were calculated by NEMD. The results are in good agreement with
static calculations and experimental measurements. 

Specific heat capacities were calculated by the direct method and from VACF. The
direct method provides results greater than experimental measurements, and with
much larger error than the VACF method. On the other hand, the VACF allowed to
analyse the phonon density of states and provide results much more accurate. The
results obtained with ClayFF are very close to previous experimental
measurements, and results from IFF are slightly smaller. The DOS obtained from
VACF are in good agreement with infrared spectroscopy measurements, and the
differences between IFF and ClayFF arise mainly from the bond description in
silicates.

Cleavage energy calculations were performed on both \ce{M1} and \ce{M3 C3S}
polymorphs, using a temperature gradient method to relax superficial ions to
configurations of lower energy. These calculations allowed the construction of
equilibrium shapes which are significantly different. The \ce{M1} crystal
possess three facets against seven for the \ce{M3} polymorph. The energies
obtained for both polymorphs are in the same range. Despite different
morphologies, low index crystal planes with the lowest energies are the (100)
and (001) for both polymorphs. The cleavage energies were calculated for a
relatively limited number of planes, and further calculation could lead to even
more accurate shapes.

\section*{Acknowledgments}
The authors acknowledge Brazilian science agencies CAPES (PDSE
process n\si{\degree}88881.188619/2018-01) and CNPq for financial support.
	
\bibliographystyle{elsarticle-num}
\biboptions{sort&compress}
\bibliography{ccr2020}

\end{document}